\documentclass[11pt,a4paper]{article}

\RequirePackage[T1]{fontenc}
\RequirePackage[utf8]{inputenc}
\RequirePackage{calc}
\RequirePackage{indentfirst}
\RequirePackage{fancyhdr}
\RequirePackage{graphicx,epstopdf}
\RequirePackage{lastpage}
\RequirePackage{ifthen}
\RequirePackage{lineno}
\RequirePackage{float}
\RequirePackage{amsmath}
\RequirePackage{amssymb} 
\RequirePackage{setspace}
\RequirePackage{enumitem}
\RequirePackage{booktabs} 
\RequirePackage{titlesec}
\RequirePackage{etoolbox} 
\RequirePackage{tabto} 
\RequirePackage{xcolor, colortbl} 
\RequirePackage{soul} 

\RequirePackage{multirow}
\RequirePackage{microtype} 
\RequirePackage{tikz} 
\RequirePackage{totcount} 
\RequirePackage{changepage} 
\RequirePackage{paracol} 
\RequirePackage{attrib} 
\RequirePackage{upgreek} 
\RequirePackage{array} 
\RequirePackage{tabularx}
\RequirePackage{pbox} 
\RequirePackage{ragged2e} 
\RequirePackage{amsthm}
\usepackage{dsfont}
\usepackage{centernot}
\usepackage{bm}
\usepackage[ruled,vlined,linesnumbered]{algorithm2e}
\usepackage{authblk}
\usepackage{fullpage}

\SetKwInput{KwUpdate}{Update}

\newcounter{theorem}
\newtheorem{Theorem}[theorem]{Theorem}

\newcounter{lemma}
\newtheorem{Lemma}[lemma]{Lemma}

\newcounter{corollary}
\newcounter{proposition}
\newcounter{remark}
\newcounter{definition}
\newcommand{\m}{\mathcal}

\DeclareMathOperator*{\argmax}{\arg\!\max}

\newcommand{\vx}{\bm{x}}
\newcommand{\vy}{\bm{y}}
\newcommand{\vz}{\bm{z}}
\newcommand{\A}{A}

\newcommand{\bbQ}{\mathbb{Q}}
\newcommand{\bbP}{\mathbb{P}}

\newcommand{\dTV}{\bm{\mathrm{d_{TV}}}}
\newcommand{\dH}{\bm{\mathrm{d_H}}}

\newcommand{\Expt}{\mathbb{E}}

\newcommand{\RR}{\mathbb{R}}
\newcommand{\eps}{\varepsilon}

\newcommand{\LDPC}{\bm{\mathrm{LDPC}}}

\newcommand{\Revision}[1]{{#1}}

\begin{document}

\title{On Compressed Sensing of Binary Signals for the Unsourced Random Access Channel }	
\date{}

\author[1]{Elad Romanov \thanks{E-mail: elad.romanov@mail.huji.ac.il}}
\author[1]{Or Ordentlich\thanks{E-mail: or.ordentlich@mail.huji.ac.il }}

\affil[1]{School of Computer Science and Engineering, The Hebrew University, 
	Jerusalem, Israel}

\maketitle	

\begin{abstract}
	Motivated by applications in unsourced random access, this paper develops a novel scheme for the problem of compressed sensing of binary signals. In this problem, the goal is to design a sensing matrix $A$ and a recovery algorithm, such that the sparse binary vector $\mathbf{x}$ can be recovered reliably from the measurements $\mathbf{y}=A\mathbf{x}+\sigma\mathbf{z}$, where $\mathbf{z}$ is additive white Gaussian noise. We propose to design $A$ as a parity check matrix of a low-density parity-check code (LDPC), and to recover $\mathbf{x}$ from the measurements $\mathbf{y}$ using a Markov chain Monte Carlo algorithm, which runs relatively fast due to the sparse structure of $A$. The performance of our scheme is comparable to state-of-the-art schemes, which use dense sensing matrices, while enjoying the advantages of using a sparse sensing matrix.	
\end{abstract}

\section{Introduction}

The emergence of the Internet of Things (IoT) has motivated much research interest in designing communication protocols for massive machine-to-machine type communication. This type of communication setup is characterized by a large number of users that transmit simultaneously to the same receiver, while each of these users has a very short message to send. In addition, since IoT sensors are often required to be extremely cheap, the transmission scheme must be as simple as possible, and the design objective is to minimize the energy-per-bit, $E_b/N_0$, under a reliability constraint.

In~\cite{polyanskiy2017perspective}, Polyanskiy defined a communication model capturing the challenges in massive machine-to-machine type communication. In this model there is an unbounded number of potential users, among which only $k$ are active at each frame. Each active user has a message of $B$ bits to transmit, and transmission takes place over a multiple access channel (MAC). Since the number of users is unbounded, the receiver cannot recover the identities of the active users (as this information has unbounded entropy, assuming all potential users are equally likely to transmit, and the channel has bounded capacity). Thus, the receiver's goal is to recover a list of $k$ messages, that contains ``most'' of the transmitted messages, without identifying the sender of each message. This setup is therefore called \emph{the unsourced random access channel}~\cite{vncc19}. 
The performance of a communication scheme over the unsourced random access channel is assessed by the tradeoff it achieves between energy-per-bit and the per-user probability of error (PUPE), which is the probability that the message transmitted by an active user did not enter the list of messages the receiver outputs.

Over the last few years, there has been great interest in developing efficient low-complexity schemes for the unsourced random access channel~\cite{ordentlich2017low,vncc19,marshakov2019polar,calderbank2020chirrup,csly19,amalladinne2020coded,gmaft19,fengler2019sparcs-isit,kafp19,amalladinne2020AMP,facenda2020efficient,decurninge2020tensor,shyianov2020massive,wgzypc20}.
A natural approach for this setup is for all users to transmit codewords from the same codebook. It can be easily seen that if $A\in\RR^{n\times 2^B}$ is a matrix whose columns are the codewords of this codebook, and $\mathbf{x}\in\{0,1\}^{2^B}$ is a vector whose $i$th entry equals $1$ if one of the active users chose message $i$ and $0$ otherwise, the channel output is $\mathbf{y}=A\mathbf{x}+\sigma\mathbf{z}$, where $\mathbf{z}$ is white Gaussian noise.\footnote{We have assumed here for simplicity that no message was chosen by more than one user.} Since the number of active users $k$ is typically of the order of tens to hundreds, and is much smaller than $2^B$, whereas the blocklength $n$ is typically on the order of $10^4$ to $10^5$, the problem of designing efficient codebooks and decoding algorithm for the unsourced random access channel corresponds to designing the sensing matrix $A$ and a recovery algorithm for a compressed sensing problem~\cite{polyanskiy2017perspective}. However, this compressed sensing problem has two non-standard features: (i) the dimensions of the problem are huge (recall that $B=100$ is a typical number); (ii) the sparse vector $\vx$ is binary, in contrast to the standard compressed sensing setup where the nonzero entries can take values in an interval within the real line.

To address the dimensions of the compressed sensing problem, Amalladinne~\emph{et al.}~\cite{amalladinne2020coded} introduced the coded compressed sensing framework, where the $B$ message bits are divided to smaller chunks, and are encoded on different sub-blocks.\footnote{\Revision{This is somewhat related to ideas that have previously appeared in the compressed sensing and group testing literature, wherein one constructs the measurements matrix by combining an ``outer'' and ``inner'' code. See e.g \cite{cormode2006combinatorial,gilbert2007one,ngo2012efficiently}.}} This idea breaks the original compressed sensing problem into a sequence of compressed sensing problems with manageable dimensions, which can be handled via existing tools from the compressed sensing literature. A difficulty that arises under this framework is that the sub-messages eventually have to be stitched to one long message, and a tree code was developed in~\cite{amalladinne2020coded} for this purpose. While there has been many important advances in the field since the first appearance of the coded compressed sensing framework~\cite{AmalladinneICASP}, the idea of first solving small compressed sensing problems and then leveraging the solutions to obtain estimates of the entire message still appears in one way or another in practically all schemes achieving state of the art performance.

Motivated by the above, the focus of this paper is the design of sensing matrices and efficient decoding algorithms for (small dimensions) compressed sensing of binary signals. Originally, Amalladinne~\emph{et al.}~\cite{amalladinne2020coded} treated this challenge by designing the sensing matrix based on BCH codes, and using off-the-shelf recovery algorithms, such as LASSO or non-negative least squares (NNLS), for decoding. 
\Revision{The main weakness of this approach is that it fails to exploit the fact that the entries are binary.}
Later, Fengler, Jung and Caire~\cite{fengler2019sparcs-isit} suggested to use Sparse regression codes with approximate message passing (AMP) decoding.
The main benefit of the AMP decoder is that it allows to incorporate any prior one has on the signal $\mathbf{x}$, and not just sparsity. Consequently, it achieves excellent performance when $\mathbf{x}$ is a binary sparse vector and the sensing matrix $A$ is i.i.d. Gaussian. This framework has more benefits, for example it allows one to efficiently jointly decode all compressed sensing problems corresponding to the different sub-blocks, and one can even iterate between the AMP decoder and the tree decoder~\cite{amalladinne2020AMP}.

In this paper, we propose an alternative design for a sensing matrix $A$ and a decoding algorithm. Our sensing matrix $A$ is taken as the parity check matrix of a low-density parity check (LDPC) code, thought of as a matrix over the reals. The decoder is based on the Markov Chain Monte Carlo (MCMC) method, more specifically, Glauber dynamics. This method performs a random walk over a Markov chain whose state space consists of all possible values of $\vx$ and whose stationary distribution is the conditional probability of $\vx$ given the measurement $\vy$. Due to the sparse structure of the matrix $A$, each step in the random walk can be simulated with a low computational cost.

For the compressed sensing problem with binary signals problem, our proposed framework achieves comparable performance to that of AMP with a Gaussian sensing matrix. However, in contrast to the AMP framework, which is based on sensing matrices that are Gaussian i.i.d., or ``Gaussian i.i.d.-like'', our sensing matrix is sparse. The sparsity of the sensing matrix $A$ in compressed sensing of binary signals has several benefits that go beyond the unsourced random access application:
\begin{itemize}
	\item \textbf{Storage.} Storing a sparse matrix requires less memory resources than storing a dense unstructured matrix, such as a matrix sampled from the i.i.d. Gaussian ensemble.\footnote{However, the AMP algorithm often works very well for compressed sensing of binary signals even when the Gaussian i.i.d. matrix $A$ is replaced with a sensing matrix that is dense yet easy to store. For example,~\cite{aprck20} suggests to take $A$ as a sub-sampled Hadamard matrix.}
	\item \textbf{Joint source-channel coding with local updates.} Consider the problem of storing a sparse binary vector $\mathbf{x}\in\{0,1\}^M$ with Hamming weight at most $k$, in an array of $n$ noisy memory cells. By noisy memory cells, we mean that the value read from memory cell $i$ is modelled as $s_i+z_i$, where $s_i$ is the stored value and $z_i$ is additive noise, say Gaussian.\footnote{\Revision{This is a reasonable model for magnetic recording (ignoring intersymbol interference)~\cite{isw98} and for flash memories (ignoring further impairments like cross talk)~\cite{yemm13}}} Note that this is actually a joint-source channel coding problem where the source is $\mathbf{x}\in\{0,1\}^M$, the channel is Gaussian and can be used $n$ times, and the distortion measure is Hamming distortion.
	It is often desirable to use update efficient schemes. In such schemes changing one bit in the input vector $\mathbf{x}$, should correspond to changing the content of a small number of memory cells. See, e.g.,~\cite{mcw14}. When the encoding scheme is $\mathbf{s}=A\mathbf{x}$, an update in one coordinate of $\mathbf{x}$, say $x_i$, corresponds to adding (removing) the $i_{th}$ column of $A$ to (from) $\mathbf{s}$. If each column has a small number of nonzero entries, the update involves changing the stored value in a small number of cells. Thus, using a matrix $A$ with sparse columns is highly desirable.
	\item \textbf{Group testing.} In group testing the goal is to detect a set of at most $k$ defective items from $M$ possible items. To this end we designate by $\mathbf{x}\in\{0,1\}^M$ the vector whose nonzero entries are defective. We have $n$ measurements of $\mathbf{x}$, each corresponding to a different "pool". Each pool is a subset of $[n]$, and the corresponding measurement is obtained by passing the number of defective items in the pool, denoted by $\ell$, through some noisy channel $P_{\mathbf{Y}|L}(\mathbf{y}|\ell)$. See, definitions 3.1 and 3.3 in~\cite{ajs19}. The typical case is that the channel depends on the number of defective items $\ell$ only through the indicator on the event $\{\ell>0\}$, but the general model allows the measurement to be distributed as $\ell+\sigma z$, where $z\sim\mathcal{N}(0,1)$. Thus, with this model the design of the group testing scheme corresponds to designing a binary sensing matrix $A\in\{0,1\}^{n \times M}$, and the measurements are $\mathbf{y}=A \mathbf{x}+ \sigma\mathbf{z}$. Using pools, corresponding to the rows of A, with small Hamming weight, results in simpler tests. For example, the original application for which the group testing framework was developed was detection of syphilis among a large group of patients, using a small number of tests. Using pools with small Hamming weight means that we need to mix samples from less patients in each pool, which results in less work for the lab technician.
\end{itemize}

In Section~\ref{sec:binary-cs} we formalize the problem of compresses sensing of binary signals, present our suggested construction for the sparse sensing matrix, and our MCMC-based recovery algorithm. Some theoretical analysis and justification for our suggested method is also given. In Section~\ref{sec:simulations}, we evaluate the performance of the proposed scheme numerically, and compare it to other state-of-the art schemes for the compressed sensing of binary signals problem. We also evaluate the performance of an end-to-end communication scheme for the unsourced random access channel with a small amount of feedback, which uses the proposed compressed sensing of binary signals scheme as an important ingredient.
Section~\ref{sec:conclusion} is devoted to conclusion and additional discussion.

\section{Compressed Sensing of Binary Signals}\label{sec:binary-cs}

We now define a formal mathematical model for the problem studied in this paper. Consider a linear inverse problem of the form 
\begin{align}
	\vy = \A\vx + \sigma{\vz},\label{eq:model}
\end{align}
where $\bm{x}\in \RR^M$ is an unknown signal, to be recovered; $\A\in \RR^{n\times M}$ is a (known) linear measurement matrix; and $\vz\in\RR^n$ is i.i.d. Gaussian noise: $z_1,\ldots,z_n \overset{i.i.d.}{\sim} \m{N}(0,1)$. This problem becomes especially interesting in the \emph{under-determined regime}, where the number of samples $n$ is smaller than the signal dimension $M$ -- here, clearly, one cannot recover $\vx$ generically, and it is necessary to make additional structural assumptions on $\vx$. In compressed sensing, one assumes that $\vx$ is a sparse vector, where the number of non-zero entries $k$ is very small compared to $M$. Perhaps the most fundamental result in sparse recovery states that, in order to recover exactly any $k$-sparse $\vx$ from noiseless measurements $\vy=\A\vx$, one in fact needs only $n=O\left(k\log (M/k)\right)$ linear measurements, where the sensing matrix $\A$ is taken to be an i.i.d. Gaussian random matrix; \Revision{see, e.g, \cite[Chapter 9]{foucart2013mathematical}}. The recovery procedure itself, while not linear, can be formulated as a convex program which is computationally easy to solve. In recent years, a vast literature on compressed sensing has formed, spanning new theory, low-complexity algorithms and new constructions of good sensing matrices, beyond the i.i.d. Gaussian setup. We make no pretense to give a literature review on this topic; for a starting point we refer primarily to surveys \cite{donoho2005sparse,candes2005decoding,candes2006near,candes2006robust,baraniuk2007compressive,duarte2011structured,elad2010sparse,eldar2012compressed,foucart2013mathematical,marques2018review}.

We consider a setting where $\vx$ is constrained to be in a discrete set, on top of being sparse. Specifically, we shall assume it is binary: $\vx\in \{0,1\}^M$. As described above, this problem is closely related to communication over the unsourced random access channel, but problems of this form have received some attention in the past; see, for example, \cite{brunel1999euclidean,thrampoulidis2019simple,reeves2019all,jkr11}. 

Throughout, we will assume a sparse binary prior for $\vx$. Specifically, let $k$ be the expected sparsity, and denote $\rho=k/M$. The coordinates of $\vx$ are assumed i.i.d. Bernoulli random variables:
\begin{align}
	x_1,\ldots,x_M \overset{i.i.d.}{\sim} \mathrm{Bernoulli}(\rho),\label{eq:Xprior}
\end{align}
that is, $\Pr(x_i=1)=\rho$ and otherwise $x_i=0$. Clearly, the expected number of non-zero entries is just $\Expt\|\vx\|_0=k$. 
A recovery algorithm for $\mathbf{x}$ from $\mathbf{y}$ is a mapping $\hat{\mathbf{x}}:\RR^n\to\{0,1\}^M$. The performance of a recovery algorithm is measured in terms of the bit error rate (BER) it attains
\begin{align}\label{eq:ber}
	\mathrm{BER}(\vx,\widehat{\vx}) = \frac{1}{k} \sum_{i=1}^M \Pr(x_i \ne \widehat{x}_i(\vy)),
\end{align}
where the probability is taken with respect to both the additive noise, as described in ~\eqref{eq:model}, and the signal prior~\eqref{eq:Xprior}. Note that the normalization in~\eqref{eq:ber} is by the expected sparsity $k$, rather than by the length $M$ of the vector $\mathbf{x}$. Since typically $k\ll M$ in compressed sensing, normalizing by $M$ would yield a very small BER for any reasonable estimator, and normalizing by $k$ therefor makes more sense.

Given a signal dimension $M$ and budget of measurements $n$, one would typically like to: (i) Construct ``good'' sensing matrices $\A$, that allow for noise-robust recovery of signals with as little sparsity (large $k$) as possible; (ii) Come up with low-complexity recovery algorithms for recovering $\vx$ from $\vy$. As for (ii), note that one would like to go beyond off-the-shelf compressed sensing algorithms, such as the LASSO \cite{hastie2015statistical,wainwright2009sharp,gamarnik2017sparse} or Non-Negative Least Squares (NNLS) \cite{donoho2005sparse,bruckstein2008uniqueness,slawski2011sparse,meinshausen2013sign,foucart2014sparse,kueng2017robust,amalladinne2020coded}, that are designed with any real or positive signal in mind, and find algorithms that explicitly leverage the binary structure of the signal, so to attain an advantage in terms of recovery performance. In this paper we address these two points: for the sensing matrix, we propose to use sparse matrices based on LDPC codes; as for the recovery algorithm, we propose to use an MCMC sampling method that approximates the optimal (in terms of bit error probability) MAP estimator. 

\subsection{Sensing matrices from LDPC codes}\label{sec:binary-cs-matrix}

We consider sensing matrices based on Gallager's ensemble of LDPC codes \cite{gallager62}. Denote by $\LDPC(\nu,s;M,n)$ the following ensemble of random bipartite and biregular graphs, described below:
\begin{itemize}
	\item One side of the graph has $M$ vertices, which we call ``variables'' (also: left side), and the other has $n$ vertices, called ``factors'' (also: right side).
	\item For simplicity, assume $\nu M = s n$. Each variable has degree $\nu$, meaning it is connected to exactly $\nu$ factors; each factor has degree $s$. \Revision{Thus, there are exactly $\nu M = s n $ edges in the graph.}
	
	\item The edges of $\m{G}\sim \LDPC(\nu,s;M,n)$ are sampled according to the following procedure. The procedure runs in $\nu$ rounds, so that in every round one introduces $M/s$ new factors,\footnote{We assume $M/s$ is integer for simplicity} by randomly partitioning the variables $[n]$ into $M/s$ parts of size $s$ each, namely,
	\[
	[n]=\bigcup_{i=1}^{M/s} S_i\,,\quad S_i\cap S_j=\emptyset,\quad |S_i|=s\quad \textrm{for all } 1 \le i, j \le n\,,i\ne j\,.
	\]
	For every new factor $1\le i \le M/s$ introduced in this round, one adds an edge between $i$ and all the variables in the corresponding $S_i$.
\end{itemize}
The sensing matrix \Revision{$\A\in \{0,1\}^{n\times M}$} is taken to be the adjacency matrix of a randomly sampled graph $\m{G}\sim \LDPC(\nu,s;M,n)$, that is,
\[
\A_{i,j} = A(\m{G})_{i,j} = \begin{cases}
	1 \quad&\textrm{there is an edge in $\m{G}$ between factor $i$ and variable $j$} \\
	0 \quad&\textrm{otherwise}
\end{cases} \,.
\]


The idea of constructing sensing matrices from bipartite graphs is not new. It is known that when $\m{G}$ is a sufficiently good expander, the corresponding adjacency matrix $A$ is a good sensing matrix; \Revision{see, for example, \cite{indyk2008near,berinde2008combining,gilbert2010sparse,jafarpour2009efficient}, \cite[Chapter 13]{foucart2013mathematical} and the references therein}. Specifically, ensembles of LDPC codes have also been considered previously for compressed sensing \cite{arora2012message,dimakis2012ldpc,zhang2012verification}. 

It is worthwhile to recall, at this point, that the recovery problem we consider here is more structured than the ``standard'' compressed sensing setup: on top of being sparse, we assume the unknown signal is \emph{binary}, and in particular \emph{non-negative}. Past results have shown that the non-negativity assumption may give a considerable advantage in terms of the required number of measurements, as well as robustness to noise; see, for example, \cite{donoho2005sparse,bruckstein2008uniqueness,khajehnejad2010sparse,slawski2011sparse,meinshausen2013sign,foucart2014sparse,kueng2017robust}. 

We would like to especially mention the results of \cite{khajehnejad2010sparse}. We say that a bipartite graph with left degree $\nu$ is an $(r,\eps)$-expander if for every set $|S|\le r$ of left vertices, one has $|N(S)|\ge (1-\eps)\nu|S|$, $N(S)$ being the neighbors of vertices in $S$. The results of \cite{khajehnejad2010sparse} state that a bipartite left-regular $(r,1-1/\nu)$-expander yields, after applying a very small perturbation to the entries of the adjacency matrix, a sensing matrix where all \textbf{non-negative} $\lceil r/\nu-1\rceil$ sparse vectors $\vx$ can be recovered from $\vy=\A\vx$ (noiseless measurements). This guarantee, for non-negative signals, is \emph{considerably} better than what one has without the nonnegativity constraint -- to get recoverability guarantees for ``general'' compressed sensing, one needs considerably larger expansion (smaller $\eps$), see for example \cite[Chapter 13]{foucart2013mathematical}. 
\Revision{
	There are well-known connections between the decodability of LDPC codes and their expansion properties \cite{richardson2008modern}. For example, for a slightly different ensemble of LDPC codes (that contains Gallager's ensemble), one can show \cite[Theorem 8.7]{richardson2008modern} (this result first appeared in~\cite{bm01}) that with high probability, a random graph is an 
	$(\alpha^* M, 1-1/\nu)$-expander, where $\alpha^*$ is the positive solution of 
	\begin{equation*}
		\frac{\nu-1}{\nu} h_2(\alpha) - \frac{1}{s}h_2\left(\alpha \frac{s}{\nu}\right) - \alpha \frac{s}{\nu}h_2\left(\frac{\nu}{s}\right) = 0 \,,
	\end{equation*}
	and $h_2(p)=-p\log(p)-(1-p)\log(1-p)$ is the binary entropy function. While not precisely applicable for our setup (which uses Gallager's ensemble), the following calculation could nonetheless be thought of as a crude heuristic. For example,
	in the setup we consider later on in the numerical experiments, corresponding to a typical use-case for detection in unsourced random access, $M=2^{14}$, $n=2^{11}$, $\nu=16$, $s=128$, one can solve the above equation numerically and get $\alpha^*\approx 0.993$. Together with \cite{khajehnejad2010sparse}, this hints that $k=\alpha^*M/'\nu \approx 101$ -sparse non-negative signals can be consistently recovered. In fact, the experiments indicate that practically, binary signals with considerably more non-zeros can be recovered reliably in this setting, see Section~\ref{sec:simulations-binary-cs}.
}

\subsection{MCMC algorithm for recovery}\label{sec:binary-glauber}

Recall that for a given sensing matrix $A$, our goal is to construct an estimator $\widehat{\vx}=\widehat{\vx}(\vy)$ such as to minimize the per-bit error rate (BER), as defined in~\eqref{eq:ber}.
Clearly, the optimal estimator in the sense of minimizing the BER is simply the per-coordinate maximum a posteriori (MAP) estimator:
\begin{equation}\label{eq:per-coordinate-MAP}
	\widehat{x}_{\mathrm{BER},i} = \argmax_{\widehat{x} \in \{0,1\}} \Pr\left(x_i=\widehat{x} \,\big|\, \vy \right) \,,\quad \textrm{ for all } 1 \le i \le M\,.
\end{equation}
Computing the posterior $\Pr(x_i\,|\,\vy)$ is a formidable task: it requires one to marginalize over all other coordinates $\ell\ne i$. From a computational point of view this is highly nontrivial, since the coupling between the coordinates of $\vx$, as induced by $\A$, creates a strong cross-coordinate dependency conditioned on $\vy$. 

We propose to mitigate this difficulty by \textbf{sampling}. 
Instead of marginalizing and and maximizing, we will sample an $\widehat{\vx} \in \{0,1\}^M$ from the full posterior, given by
\begin{equation}\label{eq:posterior}
	\begin{split}
		\Pr\left(\vx = \widehat{\vx} \,\big|\, \vy\right) = \frac{1}{Z} \exp \left\{ -\frac{1}{2\sigma^2}\|\vy-\A\widehat{\vx}\|^2 + \lambda \|\widehat{\vx}\|_1  \right\}
	\end{split} \,,
\end{equation}
where $\lambda=\log \frac{\rho}{1-\rho}$, $\rho=k/M$, and $Z$ is the partition function (normalization).\footnote{\Revision{To see the correctness of $\lambda=\log\frac{\rho}{1-\rho}$, note that the prior is $\Pr(x_i=1)=e^{\log \rho}$ and $\Pr(x_i=0)=e^{\log (1-\rho)}$. In other words: $\Pr(x_i=\hat{x}_i)=e^{\hat{x}_i\log \rho + (1-\hat{x}_i)\log (1-\rho)} = e^{\lambda \hat{x}_i + \log (1-\rho)} \propto e^{\lambda \hat{x}_i}$. }}
Taking the $i$-th coordinate of $\widehat{\vx}$, call it $\widehat{x}_i$, we will have obtained a sample from \Revision{$\Pr(x_i=\cdot|\vy)$}, the desired single-bit posterior distribution.


Intuition suggests that when $\widehat{x}_{\mathrm{BER},i}$ has small error, the estimator obtained by sampling, call it $\widehat{x}_{\mathrm{SAMP},i}$, should have small error as well. This is because, if the optimal error is small, the posterior $\Pr(x_i|\vy)$ must put most of its mass on $\widehat{x}_{\mathrm{BER},i}=\widehat{x}_{\mathrm{BER},i}(\vy)$; this in turn means that, with high probability over the sampling procedure, one should in fact get $\widehat{x}_{\mathrm{SAMP},i}=\widehat{x}_{\mathrm{BER},i}$. This reasoning is formalized in the following Lemma:

\begin{Lemma}\label{lem:MAP-vs-SAMP}
	Denote $\widehat{\vx}_{\mathrm{BER}}=\left(\widehat{x}_{\mathrm{BER},1},\ldots,\widehat{x}_{\mathrm{BER},M}\right)$, with coordinates given by Eq. (\ref{eq:per-coordinate-MAP}). 
	Let $\widehat{\vx}_{\mathrm{SAMP}} = \widehat{\vx}_{\mathrm{SAMP}}(\vy) \sim \Pr(\cdot\,|\,\vy)$ be a random sample from the posterior (\ref{eq:posterior}). Then:
	\[
	\mathrm{BER}\left( \vx,\widehat{\vx}_{\mathrm{SAMP}} \right) \le 2\cdot \mathrm{BER}\left( \vx,\widehat{\vx}_{\mathrm{BER}} \right)
	\]
	In other words, the bit error rate of 
	$\widehat{\vx}_{\mathrm{SAMP}}$ 
	is bounded by twice the optimal bit error rate, over all estimators. 
	
	Note that on the left-hand-side, the probability is taken both over the randomness in $\vx$ and the noise, as well as the sampling procedure used for constructing $\widehat{\vx}_{\mathrm{SAMP}}$.
\end{Lemma}

Several variants of Lemma~\ref{lem:MAP-vs-SAMP} have been proved in the past, see for example 
\cite{cover1967nearest,kudekar2016comparing,liu2017alpha}.
For completeness, we provide a short proof in the appendix, see Section~\ref{sec:proof-lem:MAP-vs-SAMP}.

Thus, we are left with the problem of sampling from the posterior $\Pr(\vx\,|\,\vy)$ -- doing so ``directly'' might seem, at first glance, essentially just as hard as maximizing the posterior (namely, need to go over all $2^M$ possible signal configurations). 
Markov-Chain Monte Carlo (MCMC) methods provide a strong toolbox for sampling, \emph{approximately}, from high-dimensional distributions. 
The idea is to construct an ergodic Markov chain such that (i) its stationary distribution is the desired (high-dimensional) distribution one would like to sample from, namely $\Pr(\vx\,|\,\vy)$ (ii) the chain is easy to propagate in time (e.g, its update rule is local). Having constructed such a chain, and assuming that it mixes sufficiently fast (which is often difficult to ensure), one can therefore efficiently sample from the desired distribution, up to high precision. 
For further background and discussion on MCMC, we refer to \cite[Chapter 3]{levin2017markov}. 
The use of
MCMC methods for solving inverse problems in signal processing and for decoding/detection in communication is by no means novel, see e.g.~\cite{neal2001monte,mezard2009information,hansen2009near,hassibi2010mcmc,hassibi2014optimized,bhkr18,dx05,lucka2012fast}. While both the idea of using LDPC codes as sensing matrices and the idea of using MCMC methods for decoding are not new, our innovation here is in combining the two concepts for the compressed sensing of binary signals problem. As will become evident below, the sparse structure of the sensing matrix constructed from an LDPC code significantly reduces the computational load from the MCMC decoder by reducing the computational cost of each iteration.

We propose to use the well-known \emph{Gibbs sampling} method, also known as \emph{Glauber dynamics}, which is a general-purpose recipe for sampling from high-dimensional distributions.  
Let $\bbQ(\vx)$ be a distribution over $\{0,1\}^M$ from which one wants to sample; in our case, of course, $\bbQ(\vx)=\Pr(\vx\,|\,\vy)$. We construct a chain $\vx^{(1)},\vx^{(2)},\ldots \in \{0,1\}^M$ starting from some (arbitrary) initial state $\vx^{(0)}$ according to the following transition rule. Suppose that the current state is $\vx^{(t)}$; one samples a coordinate to update at random, $i_t \sim \mathrm{Uniform}(\{1,\ldots,M\})$, so that $\vx^{(t+1)}_j = \vx^{(t)}_j$ for all $j\ne i_t$. 
As for coordinate $i_t$, it is sampled according to the conditional distribution of $\vx_{i_t}$, with all other coordinates fixed and given by $\vx^{(t)}$, that is: $x^{(t+1)}_{i_t}\sim \bbQ(x_{i_t}\,|\,\vx_{\sim i_t}=\vx^{(t)}_{\sim i_t})$ (we denote the vector of all coordindates, except for $i_t$, by $\vx_{\sim i_t}$). 

Applied to the posterior in (\ref{eq:posterior}), Glauber dynamics reads as follows:
\\~\\

\begin{algorithm}[H]
	\label{alg:glauber}
	\SetAlgoLined
	\caption{Glauber dynamics for binary compressed sensing}
	\KwIn{$T$ = number of steps to run; 
		$\vx^{(0)} \in \{0,1\}^M$ = initial state;
		$\vy \in \RR^n$ = measurements;\\
		parameters $\sigma^2>0$, $\lambda\in \RR$.
	}
	\For{$t=1,\ldots,T$}{
		$i_t \sim \mathrm{Uniform}(\{1,\ldots,M\})$; new coordinate to update \\
		
		Let\footnote{Here $\vx^{(t)}_{i_t=0},\vx^{(t)}_{i_t=1}$ stand for setting, in $\vx^{(t)}$, the $i_t$-th coordinate to $0,1$ respectively.} $q_0^{(t)} = \exp\left\{ -\frac{1}{2\sigma^2}\|\vy-\A\vx^{(t)}_{i_t=0}\|^2 \right\}$, $q_1^{(t)} = \exp\left\{ -\frac{1}{2\sigma^2}\|\vy-\A\vx^{(t)}_{i_t=1}\|^2 + \lambda \right\}$

		\KwUpdate {
			\Indp\\
			$x_{i_t}^{(t+1)}=1$ w.p. $p_1^{(t)}=\frac{q_1^{(t)}}{q_0^{(t)}+q_1^{(t)}}$; otherwise $\vx_{i_t}^{(t+1)}=0$\\
			{$\vx_{\sim i_t}^{(t+1)} = \vx_{\sim i_t}^{(t)}$}
		}
		
	}
	\Return{$\vx^{(T)}$}
\end{algorithm}

\vspace{10pt}

It is easy to see that the process $\vx^{(1)},\vx^{(2)},\ldots \in \{0,1\}^M$ is an ergodic Markov chain, and therefore has a unique stationary distribution. Furthermore, it is easy to verify that $\bbQ(\vx)$ is a stationary distribution of this chain. Thus, for $T$ sufficiently large, we have that indeed $\mathbf{x}^{(T)}$ is distributed as a random sample from $\bbQ(\vx)$. 
Note that when $A$ is a sparse LDPC matrix, each iteration of Algorithm~\ref{alg:glauber} is computationally very cheap. One can easily keep track of $\vy^{(t)} = \A\vx^{(t)}$ and $\|\vy-\vy^{(t)}\|^2$ across iterations, noting that an update to a coordinate of $\vx^{(t)}$ requires updating only $\nu$ coordinates of $\vy^{(t)}$, where $\nu$ is the degree of a variable in $\A$. Thus, the computational complexity of Algorithm~\ref{alg:glauber} is $O\left( T\nu \right)$, where a typical choice of $T$ should be $T=O(M\log M)$ \Revision{(see Lemma~\ref{lem:fast-mixing} below)}.

We can give the following guarantee for the mixing time of Glauber dynamics:

\begin{Lemma}[Fast mixing for Glauber dynamics]
	\label{lem:fast-mixing}
	Let $\vy\in\RR^n$, $\vx^{(0)} \in \{0,1\}^M$, $\sigma^2>0$ and $\lambda\in \RR$ be any parameters. 
	Denote the following distribution $\bbQ(\cdot)$ on the cube $\{0,1\}^M$ by 
	\[
	\bbQ^{(\infty)}(\vx) = \bbQ^{(\infty)}_{\vy,\sigma^2,\lambda}(\vx) = \frac{1}{Z}\exp\left\{ -\frac{1}{2\sigma^2}\|\vy-A\vx\|^2 + \lambda \|\vx\|_1 \right\}\,,
	\]
	where $Z$ is the partition function. Denote by $\bbQ^{(T)}(\cdot)=\bbQ^{(T)}_{\vy,\sigma^2,\lambda,\vx^{(0)}}(\cdot)$ the distribution of $\vx^{(T)}$, the state returned after running Algorithm~\ref{alg:glauber} for $T$ steps.
	Suppose that
	\begin{equation}\label{eq:sigma-cond}
		4\sigma^2 > \nu(s-1)\,,
	\end{equation}
	where $\nu$ and $s$ are respectively the variable and factors degrees in $A$.
	
	\Revision{Let $\epsilon>0$ let be a target precision.} Then for any $T\ge \left(\log(1/\eps)+\log(M)\right)\cdot \frac{4\sigma^2 M}{4\sigma^2-{\nu(s-1)}}= \Theta_{\eps,\sigma^2,\nu,s}(M\log M)$, one has\footnote{\Revision{Note that the big-Oh notation $O_{\alpha}(\cdot)$ simply indicates that the constants can depend on $\alpha$. Likewise for $\Omega_\alpha(\cdot)$, $\Theta_\alpha(\cdot)$. }}
	\[
	\dTV\left( \bbQ^{(T)}, \bbQ^{(\infty)} \right) \le \eps \,,
	\]
	where $\dTV(\cdot,\cdot)$ stands for total variation (statistical) distance.
\end{Lemma}
A proof is given in the appendix, see Section~\ref{sec:proof-lem:fast-mixing}. 


Note that Lemma~\ref{lem:fast-mixing} applies for \emph{any} $\vy\in \RR^n$ and $\sigma^2$, that do not necessary have anything to do with the model~\eqref{eq:model}. However, when $\vy,\sigma^2$ do correspond to measurements from \eqref{eq:model}, namely $\vy=A\vx+\sigma\vz$, Lemma~\ref{lem:fast-mixing}, combined with Lemma~\ref{lem:MAP-vs-SAMP}, allows us to bound the bit error rate of the estimator $\widehat{\vx}=\vx^{(T)}$ returned by running $T$ iterations of Glauber dynamics. Assuming condition \eqref{eq:sigma-cond} holds, Lemma~\ref{lem:fast-mixing} tell us that running $T=\frac{4(c+1)\sigma^2}{4\sigma^2-4\nu(s-1)}\cdot M\log M=O(M\log M)$ iterations of Glauber dynamics gives, with probability one (over $\vx,\vy$) an output $\vx^{(T)}$ whose law is $M^{-c}$-close to the law of $\widehat{\vx}_{SAMP}$, in total variation distance -- here $c>0$ can be taken as large as one likes. Recall that total variation distance is just
\footnote{\Revision{Maximization here is done over all bounded functions $f:\{0,1\}^M\to [0,1]$. Recall that the maximum is actually attained at the indicator fuction $f=\bm{1}_S$, where $S=\{x\,:\,\bbP(x)/\bbQ(x)\ge 1\}$.}}
$\dTV(\bbP,\bbQ)=\max_{0\le f \le 1} \left\{ \Expt_{\widehat{\vx}\sim \bbP}[ f(\widehat{\vx})] - \Expt_{\widehat{\vx}\sim \bbQ}[ f(\widehat{\vx}) ]\right\}$. Plugging $f(\widehat{\vx})=\frac{1}{k}\sum_{i=1}^M \bm{1}_{x_i\ne \widehat{x}_i}$ and noting that $f$ is nonnegative and bounded by $M/k$, we deduce
\begin{align*}
	\mathrm{BER}(\vx,\vx^{(T)}) 
	&\le \mathrm{BER}(\vx,\widehat{\vx}_{\mathrm{SAMP}}) + \frac{M}{k}\Expt_{\vx,\vy}\dTV(\bbP_{\widehat{\vx}^{(t)}},\bbP_{\widehat{\vx}_{\mathrm{SAMP}}}) \\
	&\le \mathrm{BER}(\vx,\widehat{\vx}_{\mathrm{SAMP}}) + M^{-c+1}/k \,,
\end{align*}
which, by Lemma~\ref{lem:MAP-vs-SAMP}, is bounded by twice the optimum BER, up to an inverse polynomial (in $M$) error. 

As a remark, we mention that in practice, MCMC methods are often implemented using \emph{annealing}, which in our case amounts to basically running Glauber dynamics with a noise variance $\sigma^2$ which is larger than the true noise.
This can help steer the system away from local maxima of $\bbQ$, by ``smoothing'' it out. 

We would like to emphasize that condition \eqref{eq:sigma-cond} is very pessimistic, and in practice Glauber dynamics appears to mix rapidly at substantially lower noise levels than predicted there. For example, in the setup we consider later on, $M=2^{14}$, $n=2^{11}$, $\nu=16$, $s=128$, so that the bound $\sigma^2_0 = \nu(s-1)/4 = 508$, translates in energy per transmitted bit as $E_b/N_0=\frac{\nu}{2\sigma^2\cdot \lg_2(M)\cdot}=\frac{1}{7(s-1)}\approx 0.001$, which is roughly $-29.5$ dB. This $E_b/N_0$ is very far from sufficient for reliable recovery of $\widehat{\vx}$ even when $k$ is small; see experiments in Section~\ref{sec:simulations-binary-cs}. In this regime, while indeed Lemma~\ref{lem:fast-mixing} holds in the sense that Glauber dynamics mixes fast, the error rate of the optimal estimator is too high to be of use. Thus, Lemma~\ref{lem:fast-mixing} should \textbf{not} be thought of as an accurate predictor for the performance of Glauber dynamics for binary compressed sensing. Instead, it should be though of as a ``sanity check'' -- an evidence that Glauber dynamics is a reasonable thing to do, at least in \emph{some} regime of the problem.

On the same note, we have observed that when $k$ is large, Glauber dynamics tends sometimes to get stuck at ``bad'' local maxima, even when the noise is moderate. To mitigate this, one can initialize $\vx^{(0)}$ \emph{reasonably close} to the true signal $\vx$, using an off-the-shelf compressed sensing solver like NNLS -- and then use Glauber dynamics as a \emph{refinement step}. Applying this additional step of Glauber dynamics may improve the performance \emph{substantially} -- see numerical results in Section~\ref{sec:simulations-binary-cs}. 
Of course, the result of Lemma~\ref{lem:fast-mixing} does not predict in any way this behavior; rather, it is completely agnostic to the starting location. Additionally, the bound on the mixing time there does not depend at all on $k$, which, as we have just mentioned, is crucial for the behavior of Glauber dynamics in practical regimes. A more sophisticated analysis of Glauber dynamics for compressed sensing of binary signals, that takes into account the points above, is an interesting problem, and, to the best of our judgement, highly nontrivial.

\section{Simulation results}\label{sec:simulations}

\subsection{Performance in binary compressed sensing}
\label{sec:simulations-binary-cs}

We start by demonstrating the performance of Glauber dynamics in the compressed sensing of binary signals setup of Section~\ref{sec:binary-cs}. 

We run many random recovery experiments, so to recover $\vx\in \{0,1\}^M$ from ${\vy=A\vx+\sigma\vz} \in \RR^n$. In all the experiments, we use $M=2^{J}$, $J=14$, $n=2^{11}$ and sparsity values $k\in \{50,100,200,300\}$. These parameters are representative of a typical setup for unsourced random access, see Section~\ref{sec:simulations-IoT} below. For each $k$, we vary the energy per transmitted bit, $E_b/N_0=\frac{E_m}{2\sigma^2\cdot J}$ (here $E_m$ is the average energy per transmitting user -- the energy of a column of $A$), and plot the corresponding bit error rate.

We plot the performance under the following schemes:
\begin{enumerate}
	\item The scheme of Amalladinne et al. \cite{amalladinne2020coded}: $A$ based on BCH codes, and NNLS decoder. To obtain a binary estimator from the NNLS solution, we simply assign every entry to its closest binary value (that is, according to whether it is smaller or greater than $1/2$). 
	\item $A$ given by a sparse LDPC matrix, with parameters $\nu=16$ (consequently $s=128$), under the following decoding algorithms:
	\begin{enumerate}
		\item NNLS.
		\item Glauber dynamics with initialization at $\vx^{(0)}=\bm{0}$.
		\item Glauber dynamics, with $\vx^{(0)}$ initialized at the NNLS solution.
	\end{enumerate}
	\Revision{When using Glauber dynamics, we always let it run for $T=10M\lg_2 M=10MJ$ iterations.} 
	\item $A$ a dense random i.i.d. Gaussian matrix \Revision{of mean $0$ and variance\footnote{Thus, $E_m=1$. Of course, in the experiments the noise level $\sigma$ is normalized according to the appropriate choice of $E_b/N_0$.} $1/n$}, with Approximate Message Passing (AMP) decoder. The denoiser used in AMP is the optimal denoiser for the i.i.d. Bernoulli source, essentially as proposed by Fengler et al. \cite{fengler2019sparcs}. AMP is a state-of-the-art algorithm for compressed sensing of binary signals, and is our main benchmark. For convenience, the exact implementation details of AMP are given in the appendix, see Section~\ref{sec:appendix-amp}.
\end{enumerate}

\begin{figure}
	\centering
	\includegraphics[width=0.45\linewidth]{{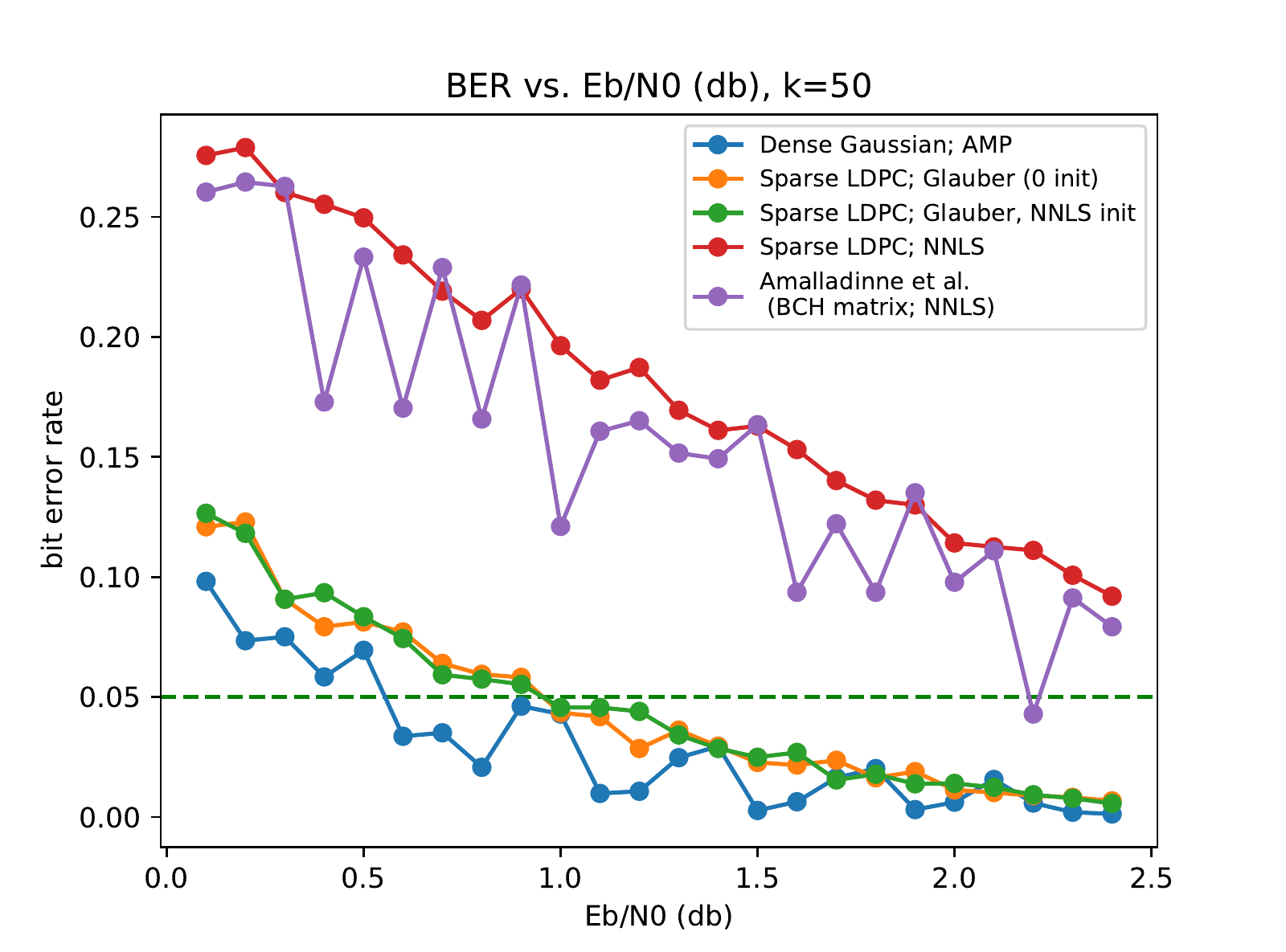}}
	\includegraphics[width=0.45\linewidth]{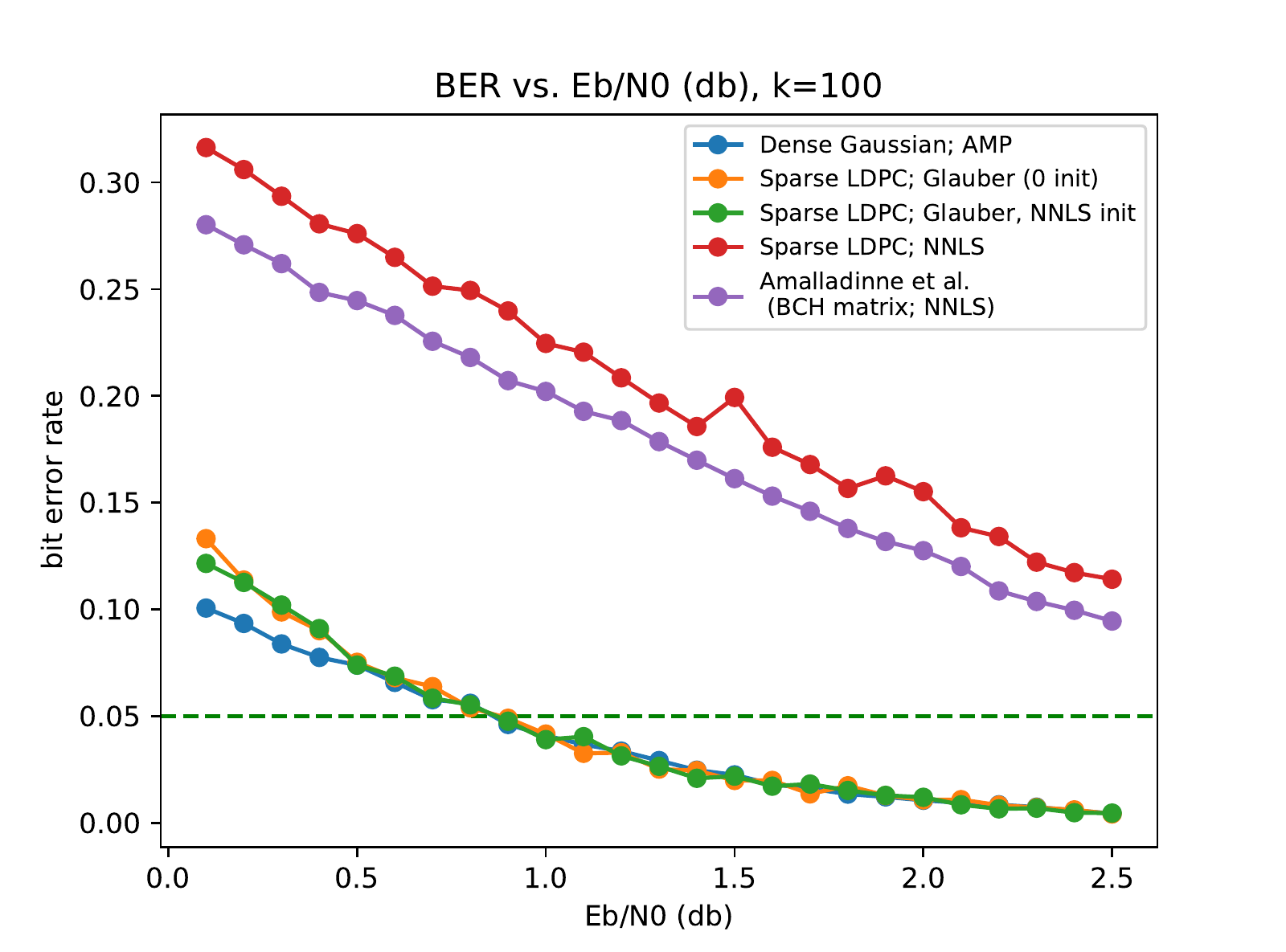}
	\includegraphics[width=0.45\linewidth]{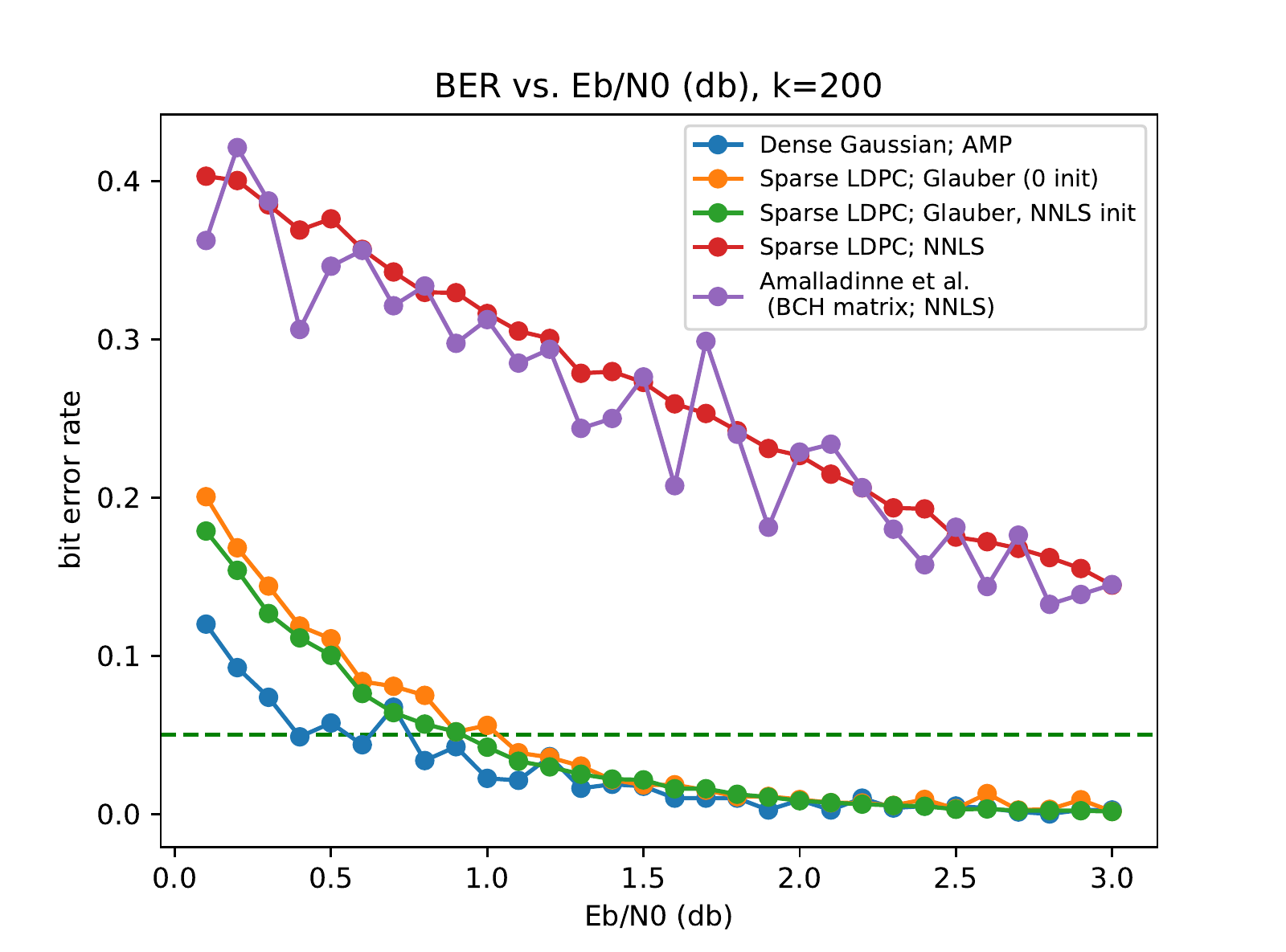}
	\includegraphics[width=0.45\linewidth]{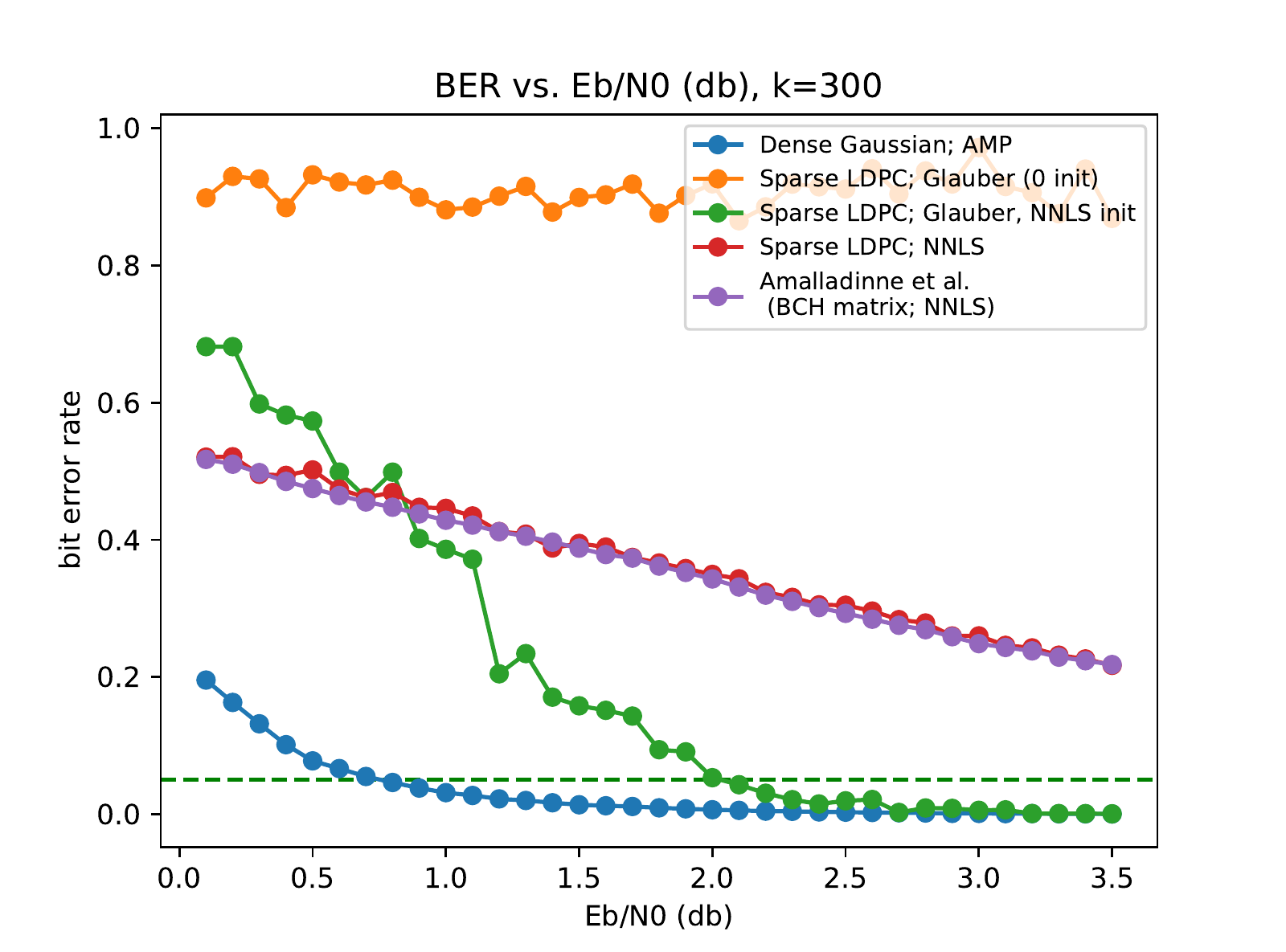}
	\caption{BER vs $E_b/N_0$ for several sparsity levels $k$. When $k$ is small to moderate, our proposal achieves state-of-the-art performance, on par with AMP on a dense Gaussian matrix. Each point on a curve is the average BER over a 100 random experiments. Dashed horizontal line: $\mathrm{BER}=0.05$.}
	\label{fig:CS-only}
\end{figure}

Our results are summarized in Figure~\ref{fig:CS-only}. We see that when the sparsity is moderate (up to $k=200$), our proposed scheme attains essentially state-of-the-art performance. However, when $k$ is large ($k=300$) performance falls short of AMP: if initialized at zero, Glauber dynamics consistently gets stuck in a local maximum, far away from the true signal; on the other hand, if one initializes Glauber dynamics with the NNLS solution, the combined scheme eventually attains performance which is substantially better than off-the-self compressed sensing solvers. 

\Revision{
	In Figure~\ref{fig:Glauber-trajectory}, we plot the evolution, across consecutive iterations, of both the BER and the ``energy'' 
	${E(\vx^{(t)})=-\frac{1}{2\sigma^2}\|\vy-A\vx^{(t)}\|^2 + \lambda\|\vx^{(t)}\|_1}$ along a single run of Glauber dynamics (initialized at $\vx^{(0)}=0$). We use $k=100$ and $E_b/N_0=1\,dB$. Note that the iterations are given in units of $MJ=M\lg_2 M$ (meaning, it is $t/MJ$). Ignoring stochastic fluctuations, we see that Glauber dynamics essentially monotonically minimizes the energy (the error, however, is not monotonically decreasing).
	
	\begin{figure}
		\centering
		\includegraphics[width=0.45\linewidth]{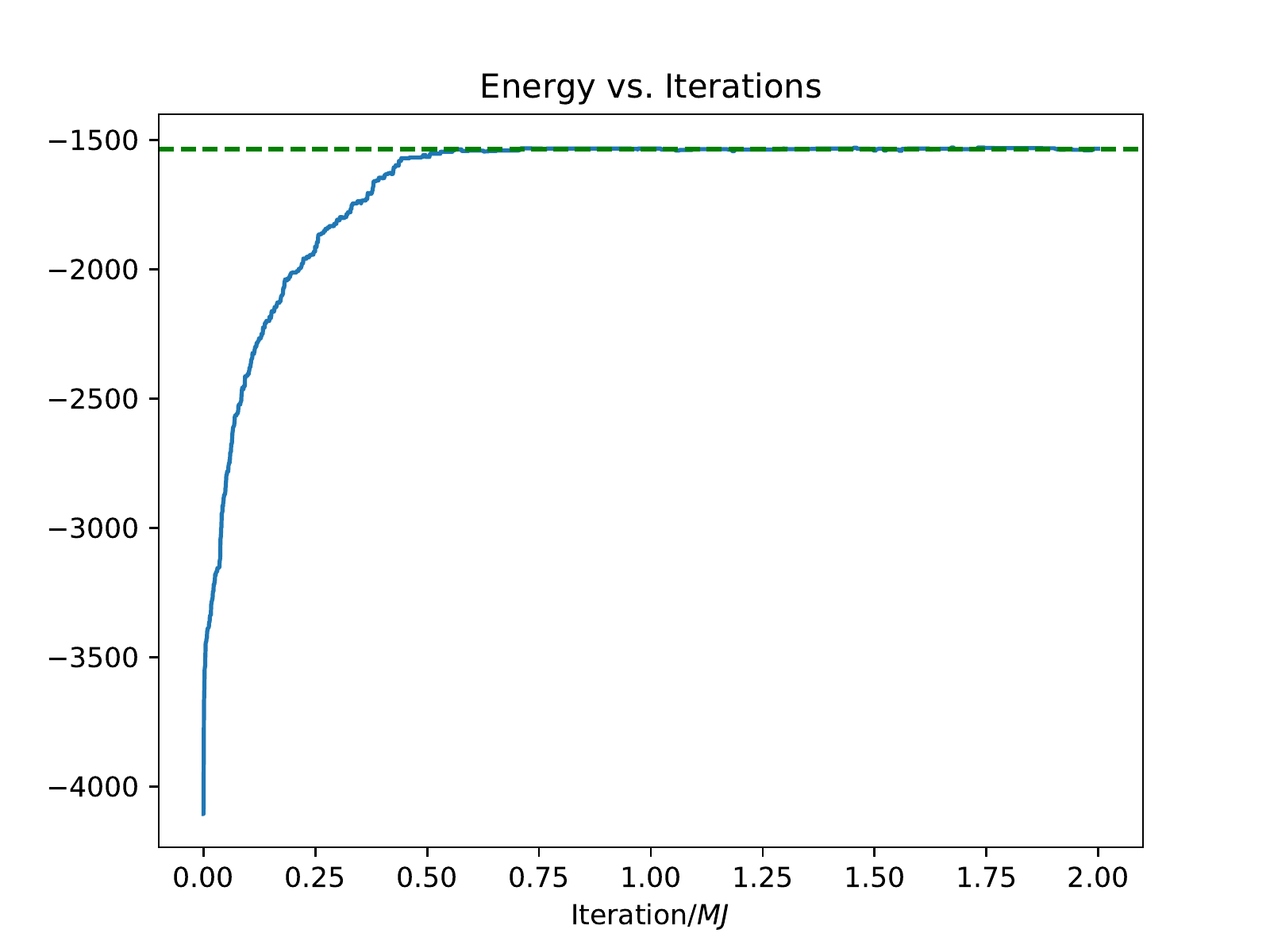}
		\includegraphics[width=0.45\linewidth]{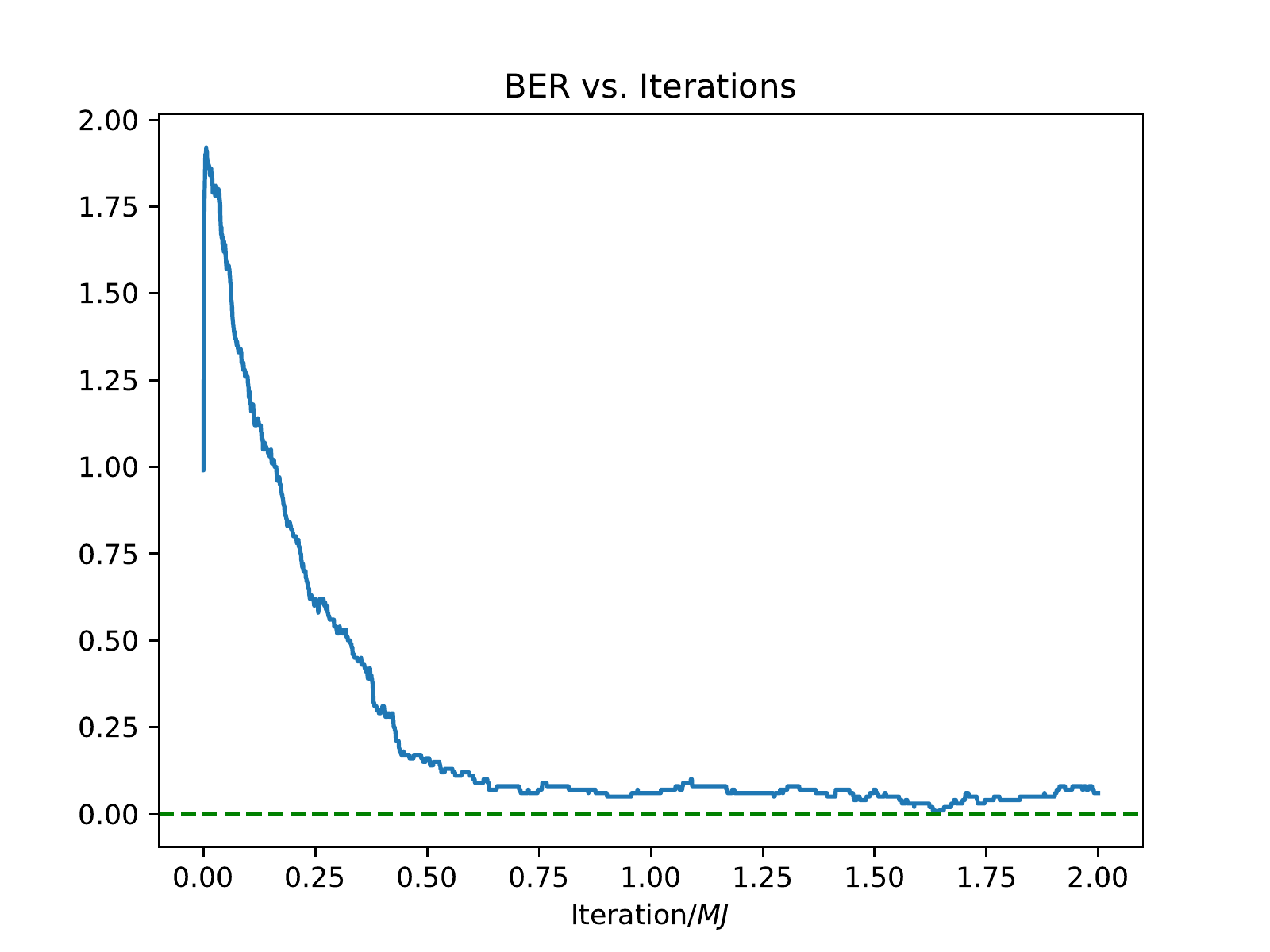}
		\caption{\Revision{Energy and error along a typical trajectory of Glauber dynamics, with $k=100$ and $E_b/N_0=1.0\,dB$. The dashed horizontal curve correspond to the energy and error respectively of the true signal $\vx$.}}
		\label{fig:Glauber-trajectory}
	\end{figure}
}

\subsection{End-to-end performance in grant-based random access}\label{sec:simulations-IoT}

As mentioned in the introduction, the compressed sensing of binary signals problem is an important component of many schemes that were proposed for communication over the unsourced random access channel. 
In this model~\cite{polyanskiy2017perspective}, communication is performed in blocks of $n$ channel uses of a Gaussian multiple access channel
\begin{align*}
	\vy=\sum_{i=1}^{K_{\text{tot}}} s_i \vx_i+\sigma \vz,
\end{align*}
where $(s_1,\ldots,s_{K_{\text{tot}}})\in\{0,1\}^{K_{\text{tot}}}$ is the ``activity pattern'' vector whose Hamming weight is $k$, $\vx_i\in\RR^n$ is the codeword transmitted by user $i$ assuming it was active, and $\vz\sim\m{N}(\mathbf{0},\mathbf{I})$ is additive white Gaussian noise (AWGN). \Revision{Note that this channel model implicitly assumes perfect power and phase control, which is often difficult to attain in practice.}
We further assume that all active users have a message of $B$ bits to transmit, and that each of these messages are independently and uniformly distributed over $[2^B]$. 
The activity pattern is assumed \emph{unknown} to the decoder, and known only locally to the transmitters, i.e., each user only knows whether or not it is active, but does not know which of the other users are active. The decoder's goal is to output a list of $k$ messages, that contain as many transmitted messages as possible. The per-user probabililty of error (PUPE) is defined as the number of transmitted messages that did not enter the list, normalized by $k$.

In this section we use the scheme we have developed above for compressed sensing of binary signals as a building block for an end-to-end communication scheme for the unsourced random access channel. We slightly deviate from the mainstream literature on unsourced random access, by allowing for some feedback to be sent from the receiver to all potential users through a broadcast channel. This option was mostly avoided until {now, with the exception of~\cite{facenda2020efficient},} as it was believed that the large number of potential users and the small payloads for each active users renders scheduling too wasteful. Recent work by Kang and Yu~\cite{KY20} establishes a connection between scheduling for the unsourced random access channel and \emph{perfect hashing} and demonstrates that in fact scheduling for the unsourced random access channel can be attained with a very small cost. Based on their observation, we propose the following scheme for the unsourced random access channel 
with an unbounded number $K_{\text{tot}}$ of potential users, among which $k$ are active users that have to send a $B$ bits message each, over $n$ channel uses:
\begin{itemize}
	\item \textbf{Phase 1:} Each active user transmits the first $J$ bits of its message over $n_1<n$ channel uses. To that end, we use a sensing matrix $A$ drawn from the $\LDPC(\nu,s;M,n_1)$ ensemble, with $M=2^J$. Each active user chooses one of the $M=2^J$ columns of $A$, corresponding to the first $J$ bits in its message, scales it by $\alpha>0$ and transmits them over the channel. Since there are $k$ active users, the channel output after $n_1$ uses is $\vy_1=\alpha A\vx+\sigma\vz$. The vector $\vx$ consists of entries in $\mathbb{Z}_+$ (all non-negative integers) and satisfies $\|\vx\|_1=k$. If all $k$ active users chose messages that begin with a different string of $J$ bits, the vector $\vx$ will further be in $\{0,1\}^M$. For our choices of $J$ and $k$ described below, typically almost all entries of $\vx$ will be binary. The basestation (which is now the receiver) applies Algorithm~\ref{alg:glauber} to estimate $\vx$. In the end, we compute $p_1^{(T+1)}(i)$ for any $i\in[M]$, and output a list consisting of the $k$ coordinates with the highest $p_1^{(T+1)}(i)$.
	\item \textbf{Phase 2:} The basestation  applies a set partitioning scheme for collision-free feedback, as described in~\cite{KY20}, for broadcasting to the users a list of the $k$ strings of $J$ prefixes it has decoded in phase $1$. Naively, this would require broadcasting a message of $k\cdot J$ bits. However, as shown in~\cite{KY20} using a more intelligent scheme, this can information theoretically be done with about $k\cdot\lg_2(e)$ bits, and practical schemes can encode this information using less than $2k$ bits. Each active user decodes the message transmitted by the basestation and finds the location of the $J$ bits prefix of its message within the list of $k$ prefixes that was transmitted.
	\item \textbf{Phase 3:} The remaining $n_2=n-n_1$ channel uses are split to $k$ slots, each of length $n'=n_2/k$. Each active user transmits the remaining $B-J$ bits of its message during the slot whose index it has decoded in phase 2. To this end, off-the-shelf point-to-point codes are used. Active users that did not find their $J$ bits prefix in the list of phase 2, do not transmit a thing in phase 3.
	
\end{itemize}

\vspace{3pt}
Note that in the end of this procedure the receiver outputs a list of at most $k$ messages. The message sent by a particular active user enters the list the decoder outputs whenever neither of the following error events occur:
\begin{enumerate}[label=(\roman*)]
	\item Another active user chose a message with the same $J$ bits prefix, causing a collision in phase 1 above.
	\item The $J$ bits prefix of the user's message did not enter the list produced by the basestation in phase 2.
	\item The user failed to decode the message sent from the basestation in phase 2.
	\item There was a decoding error in the point-to-point transmission of that user in phase 3.
\end{enumerate}
For the remainder of this discussion, we neglect the cost of phase 2 in terms of channel resources (energy and bandwidth) and its contribution to the error probability. We do this in order to avoid the need to model the broadcast channel from the basestation to the active users. In light of~\cite{KY20} the message sent by the basestation in phase 2 is significantly shorter than the messages sent by the active users. Adding this to the fact that the basestation is typically less power-constrained than the end-devices in machine-to-machine type communication, it follows that indeed phase 2 will usually have negligible effect in both aspects (bandwidth and error probability). As mentioned above, our performance figure of merit is the per-user error probability.

We conducted experiments to estimate the expected performance of this end-to-end scheme. In each experiment, each one of $k$ users generate a random message of $B$ bits to be transmitted. Let $\vx\in \{0,1,\ldots,k\}^{2^J}$ be such that $x_m=$ the number of users who sent message $m$ during phase 1. The per-user error probability for phase 1 is  
\[
\eps_1 = \frac{1}{k}\sum_{i=1}^k \Pr\left( x_{m(i)}>1 \,\vee\, m(i)\notin \m{L} \right)\,,
\]
where $\m{L}$ is the list of $k$ messages returned by the base station, and $m(i)$ is the message transmitted by user $i$. 
The error probability $\eps_1$ is estimated via Monte-Carlo simulation.
For the error of the second phase, we use the finite block normal approximation of Polyanskiy-Poor-Verd{\'u} \cite[Theorem 54]{polyanskiy2010channel}:
\begin{equation}\label{eq:finite-length}
	\frac{B-J}{n'} \approx C(P) - \sqrt{ \frac{V(P)}{n'} }Q^{-1}(\eps_2)\,,
\end{equation}
where $n'P$ is the total energy per user, $C(P)=\frac{1}{2}\lg_2(1+P)$ is the AWGN capacity and $V(P)=\frac{P(P+2)}{2(P+1)^2}(\lg_2(e))^2$ is the AWGN channel dispersion. Given a target error probability $\eps_2$, we can solve \eqref{eq:finite-length} with an equality to obtain an achievability {estimate} $P^*$ on the power $P$ necessary to attain user-basestation point-to-point error probability at most $\eps_2$. The total energy per transmitted bit (per user) is just
\[
E_b/N_0 = \frac{\frac12n'P^* + J\cdot(E_b/N_0)_{phase1}}{B}\,,
\]
where, as in the previous section, $(E_b/N_0)_{phase1}=\frac{E_m}{2\sigma^2\cdot J}$, $E_m$ being the energy of a column of $A$. For every $k$, we wanted to find the smallest $E_b/N_0$ that achieves total per-user error $\eps_1+\eps_2=0.05$. This optimization have has performed numerically.

\begin{figure}
	\centering
	\includegraphics[width=\linewidth]{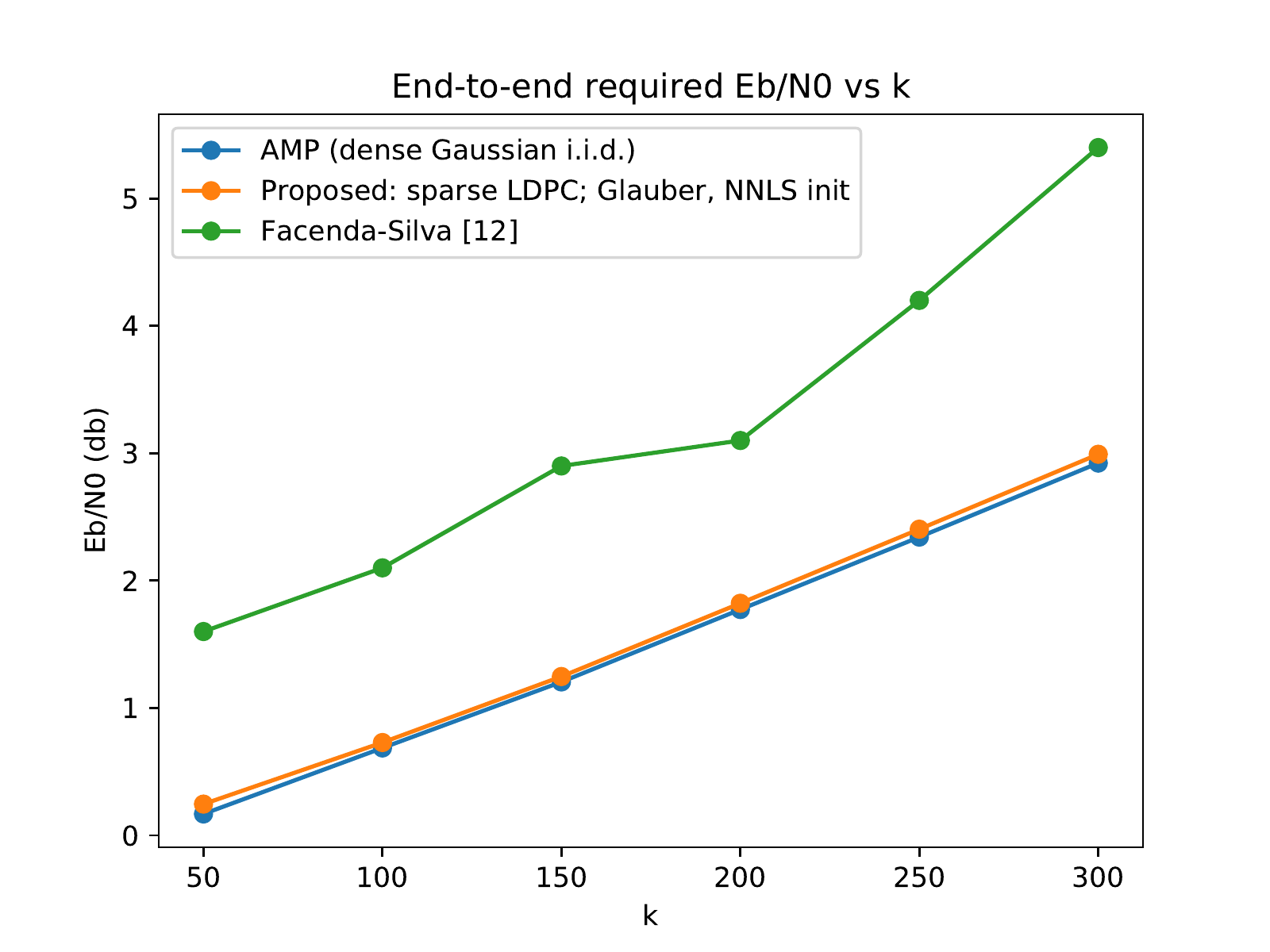}
	\caption{Total $E_b/N_0$ required to achieve end-to-end PUPE $\le 0.05$. We see that by using better compressed sensing algorithm for binary signals, significant gains can be achieved over the current state-of-the-art \cite{facenda2020efficient}. }
	\label{fig:e2e}
\end{figure}

The performance attained by this end-to-end scheme is plotted in Figure~\ref{fig:e2e}. We plot the performance corresponding to phase 1 implemented by the sensing matrix and recovery algorithm introduced in this paper, as well as an i.i.d. Gaussian sensing matrix and AMP recovery. Both implementations for phase 1 correspond to similar performance, with slight preference for the latter, and substantially improve the state-of-the-art for unsourced random access with (a small amount of) feedback~\cite{facenda2020efficient}.

\section{Conclusion and additional discussion}
\label{sec:conclusion}

\Revision{
	We have proposed a scheme for compressed sensing of binary signals, consisting of a sparse sensing matrix, based on Gallager's ensemble of LDPC codes, and a decoder based on MCMC. When used as a building block in grant-based random access, the scheme is demonstrated numerically to attain essentially state-of-the-art performance.
	To conclude, we mention several points that rise up as follow-up questions to our results.
	
	\textbf{Belief Propagation.} One of the most popular algorithms for decoding LDPC codes is Belief Propagation (BP), see e.g. \cite{richardson2008modern,mezard2009information}. 
	We have conducted very limited experiments with sum-product and max-product BP (not reported in this paper); our preliminary findings suggest that our MCMC decoder outperforms BP (in terms of its tolerance to noise), at least in the regime considered in Section~\ref{sec:simulations-binary-cs}. 
	A possible explanation for this could be that the sensing matrix $A$ has many small cycles, which severely violates the tree assumption, which is common in BP analysis of LDPC codes. A thorough study of BP for compressed sensing with binary signals is left as an interesting direction for future research.
	
	\textbf{Grantless unsourced random access.} 
	In Section~\ref{sec:simulations-IoT} we have demonstrated that our scheme can attain essentially state-of-the-art performance in grant-based unsourced random access (wherein a compressed sensing problem is solved in the first, scheduling, step). However, most previous works on unsourced random access have considered a different approach, that does not allow for feedback. The idea is to divide transmission into several blocks, and perform coding in two steps: 1) an outer code, to allow the decoder to relate (``stitch'') messages across different blocks to one another, and 2) an inner code, wherein each user codes its message (payload + parity bits) over an AWGN multiple access channel -- in this framework, decoding the inner code boils down to solving a compressed sensing problem with a binary signal. An interesting question is whether our proposed scheme can provide any gains if used to construct an inner code in this framework. In \cite{amalladinne2020AMP}, the authors propose to use a certain tree code (outer code) and an i.i.d. Gaussian sensing matrix for the inner code, together with a certain AMP decoder, that in decoding iteratively passes information between the inner and outer codes. We have tried replacing the AMP decoder with our scheme. Specifically, we considered an iterative procedure that alternates between the following steps: 1) run Glauber dynamics on each block, producing a soft decision rule for the (sparse, per-block) activity pattern; 2) a tree code inference step, that takes the per-block ``likelihoods'' produced by Glauber dynamics, and computes a posterior over the entire activity pattern, by integrating information across all the blocks; the next time we decode the inner code, this posterior is used for the new prior of the signal. Our preliminary experiments indicate that the performance of this combined scheme is rather disappointing and quite far off from state of the art \cite{amalladinne2020AMP}.
	
	\textbf{Generalizing to non-equal channel gains.}
	When discussing random access, we have modeled the received signal at the base station as $\vy = A\vx + \sigma \vz$ where ${\vx\in \{0,1\}^M}$ is the pattern of active users and $\sigma\vz$ is Gaussian noise; namely, the channel between the users and the basestation is an AWGN multiple access channel where all gains are equal. This model is based on the assumption of perfect power and phase control, which is not always realistic, and designing communication schemes for the fading model, where channel gains are not assumed equal, is desired. Generalizing our MCMC decoder to incorporate fading looks somewhat challenging. Consider a model $\vy=AH\vx+\sigma\vz$ where $H=\mathrm{diag}(h_1,\ldots,h_m)$ is a diagonal matrix of (random) fading coefficients. We would like to sample from the posterior of $\vx$ given $\vy$:
	\[
	\Pr(\vx=\widehat{\vx}|\vy) \propto \Expt_{H} \exp\left\{ -\frac{1}{2\sigma^2}\|\vy-AH\widehat{\vx}\|^2 + \lambda\|\widehat{\vx}\|_1\right\} \,,
	\]
	where notice that we now need to marginalize over $H=\mathrm{diag}(h_1,\ldots,h_M)$. This marginalization appears to complicate things considerably: in particular, in contrast to the case where $H$ is the identity matrix, in the general case it is not so straightforward to sample $x_i$ conditioned on all other coordinates. Devising an MCMC decoder that can handle fading is an interesting problem for future research.
	
}

\paragraph{Acknowledgements.}
We thank Vamsi K. Amalladinne, Jean-Francois Chamberland and Krishna R. Narayanan for valuable discussions and for kindly sharing with us their code for the scheme described in reference~\cite{amalladinne2020AMP}, and Uri Erez for valuable discussions. This work was supported in part by ISF under Grant 1791/17 and in part by the GENESIS Consortium via the Israel Ministry of Economy and Industry. The work of Elad Romanov was supported in part by an Einstein-Kaye fellowship from the Hebrew University of Jerusalem.

\appendix

\section{Omitted proofs}

\subsection{Proof of Lemma~\ref{lem:MAP-vs-SAMP}}
\label{sec:proof-lem:MAP-vs-SAMP}

Clearly, it suffices to show that $\Pr(\widehat{x}_{\mathrm{SAMP},i} \ne x_i) \le 2\Pr(\widehat{x}_{\mathrm{BER},i} \ne x_i)$ for all $i$.
Also recall that the $i$th coordinate, $\widehat{x}_{\mathrm{SAMP},i}$, is simply sampled from the posterior $\Pr(x_i|\vy)$.

Note that 
$\{\widehat{x}_{\mathrm{SAMP},i} = x_i\}\supset \{\widehat{x}_{\mathrm{SAMP},i} = \widehat{x}_{\mathrm{BER},i} \, \wedge \, \widehat{x}_{\mathrm{BER},i} = x_i\}$
, and therefore
\begin{align*}
	\Pr\left( \widehat{x}_{\mathrm{SAMP},i} \ne x_i \right) 
	&\le \Pr\left( \widehat{x}_{\mathrm{SAMP},i} \ne \widehat{x}_{\mathrm{BER},i} \, \vee \, \widehat{x}_{\mathrm{BER},i} \ne x_i \right) \\
	&\le \Pr\left( \widehat{x}_{\mathrm{SAMP},i} \ne \widehat{x}_{\mathrm{BER},i} \right) + \Pr\left( \widehat{x}_{\mathrm{BER},i} \ne x_i \right) \,.    
\end{align*}
Thus, we are done once we show that $\Pr\left( \widehat{x}_{\mathrm{SAMP},i} \ne \widehat{x}_{\mathrm{BER},i} \right) = \Pr\left( \widehat{x}_{\mathrm{BER},i} \ne x_i \right)$. By definition, for any $x_i'$ which is deterministic given $\vy$,
\begin{align*}
	\Pr\left( \widehat{x}_{\mathrm{SAMP},i} \ne x_i'\,\big|\,\vy\right) 
	&= \Pr\left(x_i\ne x_i'\,\big|\,\vy\right)    \,,
\end{align*}
where, on the left, probability is taken only with respect to the sampling procedure.
Choosing $x_i'=\widehat{x}_{\mathrm{BER},i}$, and taking the expectation over $\vy$,
\begin{align*}
	\Pr\left( \widehat{x}_{\mathrm{SAMP},i} \ne \widehat{x}_{\mathrm{BER},i}\right) 
	&= \Expt\left[ \Pr\left( \widehat{x}_{\mathrm{SAMP},i} \ne \widehat{x}_{\mathrm{BER},i}\,\big|\,\vy\right)   \right] \\
	&= \Expt\left[ \Pr\left(x_i\ne \widehat{x}_{\mathrm{BER},i}\,\big|\,\vy\right) \right] \\
	&= \Pr\left(x_i\ne \widehat{x}_{\mathrm{BER},i}\right)\,.
\end{align*}

\subsection{Proof of Lemma~\ref{lem:fast-mixing}}
\label{sec:proof-lem:fast-mixing}

The proof uses the path coupling method, which is a fundamental technique in the theory of Markov chains \cite{levin2017markov}.
\\~\\

Before getting to the proof of Lemma~\ref{lem:fast-mixing}, let us start by recalling some useful notions and set some notation.
\begin{itemize}
	\item {\bf Distance and neighbors on the hypercube}: Denote by $\m{X}=\{0,1\}^M$ the $M$-dimensional hypercube. $\m{X}$ has a natural graph structure: two vertices $\vx,\vx'\in \m{X}$ are neighbors, denoted $\vx\sim \vx'$, iff they differ in $1$ coordinate exactly. Denote by $\dH(\cdot,\cdot) : \m{X}\times \m{X}\to [0,\infty)$ the Hamming distance:
	\[
	\dH(\vx,\vx') = \sum_{i=1}^M \bm{1}_{x_i \ne x_i' } \,.
	\]
	Of course, Hamming distance coincides with the shortest path distance with respect to the graph structure on $\m{X}$.
	
	\item {\bf Coupling:} Let $X$ and $X'$ be two random variables taking values on $\m{X}$. Denote by $\bbP_X$ and $\bbP_{X'}$ the laws of $X,X'$ respectively. A coupling between $X,X'$ is a probability distribution $\bbP_{X,X'}$ on $\m{X}\times \m{X}$, whose $X$-marginal is $\bbP_X$ and $X'$-marginal is $\bbP_{X'}$. In other words, a coupling is an embedding of two random variables onto a joint probability space, defined by a joint law.  
	
\end{itemize}

\vspace{6pt}

{\bf Notation.} For $\vx\in \m{X}$, let $X_{\vx}$ be the $\m{X}$-valued random variable whose law corresponds to running one step of Glauber dynamics, starting from the initial state $\vx$ (using the notations of Algorithm~\ref{alg:glauber}, one has $\vx^{(0)}=\vx$ and so $X_{\vx} \overset{d}{=} \vx^{(1)}$, \Revision{where $\overset{d}{=}$ indicates equality in distribution}). 

The following result follows from \cite[Corollary 14.7]{levin2017markov}:

\begin{Theorem}\label{thm:path-coupling}
	
	Suppose that there is $0\le\eta<1$ with the following property: for any two neighbors $\vx\sim\vx'$, there exists a \Revision{contracting} coupling of $X_{\vx}$ and $X_{\vx'}$ \Revision{with}
	\[
	\Expt \left[ \dH\left( X_{\vx}, X_{\vx'} \right) \right] \le \eta \,.
	\]
	Then, for any initial state $\vx^{(0)}=\vx$ and $t\ge 1$, one has
	\[
	\dTV\left( \bbQ^{(t)}, \bbQ^{(\infty)}\right) \le M\cdot \eta^t \,.
	\]
	Here $\bbQ^{(t)}$ is the law of $\vx^{(t)}$, the state of Glauber dynamics at time $t$, starting from $\vx^{0}=\vx$, and $\bbQ^{(\infty)}$ is the stationary distribution.
\end{Theorem}

The proof of Lemma~\ref{lem:fast-mixing} will follow by constructing a contracting coupling between $X_{\vx}$ and $X_{\vx'}$ for any $\vx\sim \vx'$, and applying Theorem~\ref{thm:path-coupling}. The construction proceeds as follows. Let $\bm{i}\sim \mathrm{Uniform}([M])$ be a random coordinate to update. For all $\ell\in [M]$, let 
\begin{align*}
	&q_0(\ell) = \exp\left\{ -\frac{1}{2\sigma^2}\|\vy-A\vx_{\ell=0}\|^2  \right\},
	\quad q_1({\ell}) = \exp\left\{ -\frac{1}{2\sigma^2}\|\vy-A\vx_{\ell=1}\|^2 + \lambda \right\},
\end{align*}
and
\begin{align*}
	p_1(\ell) = \frac{q_1(\ell)}{q_0(\ell)+q_1(\ell)} = \varphi\left( \lambda + \frac{1}{2\sigma^2}\left\{\|\vy-\A\vx_{\ell=0}\|^2-\|\vy-\A\vx_{\ell=1}\|^2\right\} \right)\,,
\end{align*}
where $\varphi(x)=1/(1+e^{-x})$ is the logistic function.
Let ${p_1(\ell)'}$ be defined similarly, with $\vx$ replace by $\vx'$.  To obtain $X_{\vx}$ and $X_{\vx'}$, sample $U\sim \mathrm{Uniform}[0,1]$, independent of $\bm{i}$. $X_{\vx},X_{\vx'}$ coincide with $\vx,\vx'$ respectively on all coordinate $\ell \ne \bm{i}$; as for the $\bm{i}$th coordinate, \Revision{set $(X_{\vx})_{\bm{i}}=1$ if $U\le p_1(\bm{i})$ (and otherwise set to $0$), and likewise set $(X_{\vx'})_{\bm{i}}=1$ if $U\le p_1(\bm{i})'$.} Clearly, the random variables $(X_{\vx},X_{\vx'})$ constructed in this manner have the ``correct'' marginal distribution; thus, we have defined a legitimate coupling. 

It remains to show that this coupling is contracting. Since $\vx\sim \vx'$, there is a unique coordinate on which they differ, call it $\ell_0$. Observe that conditioned on $\bm{i}=\ell_0$, we have $X_{\vx}=X_{\vx'}$ exactly. On the other hand, when \Revision{$\bm{i}\ne \ell_0$}, the Hamming distance either stays the same or increases by $1$, depending on $U$. Indeed, the distance increases if and only if $\min\{p_1(\bm{i}),p_1(\bm{i})'\}< U\le \max\{p_1(\bm{i}),p_1(\bm{i})'\}$, and, conditioned on $\bm{i}$, this happens with probability $\left| p_1(\bm{i})-p_1(\bm{i})' \right|$. Thus,
\[
\Expt\left[ \dH(X_{\vx},X_{\vx'}) \right] = 1-\frac{1}{M} + \frac{1}{M}\sum_{\ell\in [M],\,\ell\ne \ell_0}\left| p_1(\ell)-p_1(\ell)' \right| \,.
\]
It remains to bound the expression on the right. 
For a variable $\ell\in [M]$, let $F(\ell)\subset [n]$ be all the factors to which it is connected in $A$; similarly, for a factor $f\in [n]$, let $V(f)\subset [M]$ be all the variables to which it is connected. Now,
\begin{align*}
	p_1(\ell) 
	&= \varphi\left( \lambda + \frac{1}{2\sigma^2}\left\{ \|\vy-A\vx_{\ell=0}\|^2 - \|\vy-A\vx_{\ell=1}\|^2  \right\} \right)    \\
	&= \varphi\left( \lambda + \frac{1}{2\sigma^2}\sum_{f\in F(\ell)} \left\{\left(y_f - \sum_{k\in V(f)\setminus \{\ell\}}x_k  \right)^2 - \left((y_f-1) - \sum_{k\in V(f)\setminus \{{\ell}\}}x_k  \right)^2\right\} \right) \\
	&= \varphi\left( \lambda + \frac{1}{\sigma^2}\sum_{f\in F(\ell)} \left(y_f -\frac12 - \sum_{k\in V(f)\setminus \{\ell\}}x_k  \right) \right)\,,
\end{align*}
and a similar expression holds for $p_1(\ell)'$, with $\vx$ replaced by $\vx'$. Since $\varphi(\cdot)$ is $1/4$-Lipschitz, 
\begin{align*}
	\sum_{\ell\in[M]\setminus\{\ell_0\}} \left| p_1(\ell)-p_1(\ell)' \right| 
	&\le \frac{1}{4\sigma^2}\sum_{\ell\in [M]\setminus\{\ell_0\}} \left| \sum_{f\in F(\ell)} \sum_{k\in V(f)\setminus\{\ell\}}     (x_k-x_k')\right|  \\
	&= \frac{1}{4\sigma^2}\sum_{\ell\in [M]\setminus\{\ell_0\}} \sum_{f\in F(\ell)} \bm{1}_{\ell_0\in V(f)} \,,
\end{align*}
where we used that $\vx$ and $\vx'$ differ only on $\ell_0$. Observe that the double sum simply counts the number of pairs $(\ell,f)$ such that $\ell_0$ and $\ell\ne \ell_0$ are both connected to the factor $f$. Recalling that the degree of all variables is $\nu$ and the degree of all factors is $s$, this is just $\nu(s-1)$. We conclude that
\[
\Expt\left[ \dH(X_{\vx},X_{\vx'}) \right] \le 1-\frac{1}{M} \left[ 1- \frac{\nu(s-1)}{4\sigma^2}\right]
\]
which is $<1$ whenever $4\sigma^2 > \nu(s-1)$
; this is exactly the condition \eqref{eq:sigma-cond}, appearing in the statement of Lemma~\ref{lem:fast-mixing}.
Applying Theorem~\ref{thm:path-coupling},  
\[
\dTV\left( \bbQ^{(t)}, \bbQ^{(\infty)}\right) \le M\cdot \left( 1-\frac{1}{M} \left[ 1- \frac{\nu(s-1)}{4\sigma^2}\right] \right)^t \le e^{-\frac{t}{4\sigma^2M} (4\sigma^2-\nu(s-1)) + \log M} \,.
\]
This bound is $\le \eps$ whenever $t$ is exceeds the quantity in Lemma~\ref{lem:fast-mixing}.

\section{Approximate Message Passing (AMP)}\label{sec:appendix-amp}

In this section we provide implementation details for the AMP algorithm used in the numeric comparisons of Section~\ref{sec:simulations}. 

Introduced by Donoho et al. \cite{donoho2009message}, AMP is a state-of-the-art recovery algorithm for solving linear inverse problems $\vy=A\vx+\vz$. Originally developed for sparse vector recovery (compressed sensing), AMP and its extensions have been shown to yield state-of-the-art algorithms for several other linear inverse problems, with Gaussian or ``Gaussian-like'' (for example, orthogonally invariant, or with some spatially coupled structure) random sensing matrices. For a very partial, selective list of references, see for example \cite{donoho2009message,donoho2013accurate,bayati2011dynamics,krzakala2012statistical,rangan2011generalized,donoho2013information,metzler2016denoising,romanov2018near,schniter2014compressive,parker2014bilinear,rush2017capacity,berthier2020state}. In unsourced random access, AMP was
first used by \cite{fengler2019sparcs-isit} and extended in \cite{amalladinne2020AMP}.

The AMP algorithm of \cite{donoho2009message} uses an iteration of the following form, starting from $\bm{x}^{(0)}=\bm{0}$, $\bm{r}^{(0)}=\bm{0}$: 
\begin{equation}\label{eq:amp}
	\begin{split}
		&\bm{b}^{(t)} = A^\top \bm{r}^{(t)} + \bm{x}^{(t)} \\
		&\bm{x}^{(t+1)} = f_t\left( \bm{b}^{(t)} \right) \\
		&\bm{r}^{(t+1)} = \vy - A\bm{x}^{(t+1)} + \frac{1}{n}\sum_{i=1}^n f_t'\left( \bm{b}^{(t)} \right) \,.
	\end{split}
\end{equation}
Here $f_t : \RR\to\RR$ is a sequence of univariate functions.\footnote{For a vector $\bm{b}$, $f(\bm{b})$ stands for applying $f$ separately to each coordinate.} The main idea of AMP (``State Evolution'') is that in the large-dimensional limit (and under certain technical assumptions), the iterates $\bm{b}^{(t)}$ behave like $\bm{b}^{(t)} \approx \vx + \sigma_t \m{N}(0,I)$; that is, additively corrupted measurements of the true signal $\vx$. The variance $\sigma_t^2$ can be estimated from $\bm{r}^{(t)}$, which approximately has the law $\m{N}(0,\sigma_t^2 I)$; conversely, it can be tracked via an explicit recursive formula. For our experiments, we use the proposal of \cite{donoho2013accurate}, and use the robust estimator ${\hat{\sigma}_t = \mathrm{median}(|\bm{r}^{(t)}|)/\Phi^{-1}(0.75)}$, where $\Phi^{-1}(\cdot)$ is the inverse Gaussian CDF. 

The details of the AMP algorithm \eqref{eq:amp} rely on the specific choice of functions $f_t$. In compressed sensing of binary signals \eqref{eq:model}, a natural choice is the MSE-optimal estimator for $\vx$ from $\bm{b}=\vx+\sigma_t \m{N}(0,I)$, namely,
\begin{align*}
	f_t\left( \bm{b} \right)_i
	\Revision{
		= \Expt\left[x_i \,\big| \,  b_i \right] 
	}
	= \frac{\frac{k}{M} \cdot e^{-\frac1{2\sigma_t^2}(b_i-1)^2}}{\left(1-\frac{k}{M}\cdot \right)e^{-\frac1{2\sigma_t^2}b_i^2}+\frac{k}{M}\cdot e^{-\frac1{2\sigma_t^2}(b_i-1)^2}} \,.
\end{align*}
This is what we use for the experiments in Section~\ref{sec:simulations}.

\bibliographystyle{ieeetr}
\bibliography{refs}

\end{document}